\documentclass[aps,prx,superscriptaddress,twocolumn,amssymb,amsmath]{revtex4-1}
\usepackage{graphicx}
\usepackage{dcolumn}
\usepackage{bm}
\usepackage{braket}
\usepackage[dvipsnames]{xcolor}
\usepackage{comment}
\usepackage{tikz}
\usepackage{amsthm}
\usepackage{mathtools}
\usepackage{upgreek}
\usepackage{placeins}

\newcommand{\f}{\ket{f}}

\newcommand{\eqnl}[2]{\begin{equation}\label{#1}#2\end{equation}}
\newcommand{\eqn}[1]{\begin{equation}#1\end{equation}}

\theoremstyle{remark}

\definecolor{myred}{RGB}{236, 17, 0}
\definecolor{myblue}{RGB}{10, 88, 153}
\definecolor{mygreen}{RGB}{26, 152, 81}
\definecolor{myorange}{RGB}{236, 137, 0}
\definecolor{LightGray}{RGB}{220,220,220}
\definecolor{LinkColor}{RGB}{167,20,49}

\usepackage[colorlinks,citecolor=blue]{hyperref} 
\usepackage{mathptmx} 

\usepackage[framemethod=tikz]{mdframed}
\newmdenv[
  linecolor=black!60,
  linewidth=0.4pt,
  roundcorner=3pt,
  tikzsetting={dash pattern=on 3pt off 2pt},
  skipabove=6pt,
  skipbelow=6pt
]{dashedbox}

\usepackage{enumitem}
\usepackage[most]{tcolorbox}
\tcbset{
  mybox/.style={
    colback=white,
    colframe=black!60,
    boxrule=0.4pt,
    sharp corners,
    frame style={dashed},
    borderline={0.4pt}{0pt}{dashed}
    before skip=6pt, after skip=6pt
  }
}

\usepackage{float}

\begin{document}

\preprint{APS/123-QED}

\title{Arbitrary-phase CZ gate and remote readout across spatially separated superconducting nodes}

\author{Anatoly Kulikov}
\email{akulikov@phys.ethz.ch}
    \affiliation{Department of Physics, ETH Zurich, 8093 Zurich, Switzerland}
    \affiliation{Quantum Center, ETH Zurich, 8093 Zurich, Switzerland}

\author{Josua D. Schär}
    \affiliation{Department of Physics, ETH Zurich, 8093 Zurich, Switzerland}
    \affiliation{Quantum Center, ETH Zurich, 8093 Zurich, Switzerland}

\author{Alonso Hernández-Antón}
    \affiliation{Department of Physics, ETH Zurich, 8093 Zurich, Switzerland}
    \affiliation{Quantum Center, ETH Zurich, 8093 Zurich, Switzerland}

\author{Aleksandr Grigorev}
    \affiliation{Department of Physics, ETH Zurich, 8093 Zurich, Switzerland}
    \affiliation{Quantum Center, ETH Zurich, 8093 Zurich, Switzerland}

\date{\today}

\begin{abstract}

Establishing a complete set of gates across spatially separated quantum nodes is fundamental for distributed quantum computation and communication networks. One of the approaches to do so consists of employing a superconducting device as a universal quantum node that is capable of storing and processing quantum information, and traveling microwave photons as means of communication.
In this work, we realize an entangling gate across a primitive two-node superconducting circuit QED network consisting of two dilution cryostats, A and B, connected via a 30-m-long cryogenic quantum link. We perform a bidirectional photon transfer protocol, where a microwave photon emitted by the communication qubit at A acquires a conditional phase based on the state of qubit B. We vary the communication photon frequency, hence tuning the conditional phase acquired during the reflection off the remote node, effectively implementing an arbitrary-phase CZ gate.
We use this entangling operation to perform single-shot readout of the qubit in node B from node A.
An interesting feature of our scheme is that it does not require precise synchronization of the active node with the remote node; in particular, remote readout can be performed without any active participation by the remote node. These protocols add to the quantum communication toolbox with superconducting circuits and provide a way of realizing basic distributed quantum computing operations.

\end{abstract}

\pacs{Valid PACS appear here}

\maketitle

\section{Introduction}

Strong interaction between individual propagating photons and stationary quantum emitters is a powerful tool in quantum optics, leading to a broad range of applications, from deterministic quantum gates \cite{Reiserer2014} and single-photon switches \cite{Chen2013b} to the realization of quantum networks \cite{Kimble2008} and chiral light–matter interfaces \cite{Lodahl2015}. In the optical domain, single-photon sources and detectors form the backbone of quantum technology, enabling quantum repeaters for long-distance communication \cite{Sangouard2011} and underpinning the efforts to integrate quantum emitters into photonic circuits for scalable quantum information processing \cite{Elshaari2020}.

Similarly, in the microwave domain, strong photon–matter interaction has been used to realize quantum non-demolition detection of itinerant microwave photons \cite{Kono2018,Besse2018},
quantum state transfer and entanglement distribution between spatially separated superconducting qubits \cite{Kurpiers2018}. Microwave single photons have also been proposed as sensitive probes of remote quantum systems, with applications ranging from quantum radar and quantum illumination \cite{Lloyd2008,Barzanjeh2015} to the spectroscopy of mesoscopic materials and strongly correlated quantum phases \cite{Bienfait2016}. Although quantum communication and photon-based detection
have been demonstrated individually, combining them in a single experiment enables a new functionality -- a deterministic remote entangling gate that does not require classical communication \cite{Penas2022}, enabling a complete gate set for distributed quantum computing with stationary superconducting qubits \cite{Barrett2005}.

Here, we take this step by leveraging the bidirectional photon transfer capability of our experimental communication setup. We employ a microwave photon as both a probe and an information carrier to a superconducting qubit located in a spatially separate computational node 30~m away from the active node. By unifying shaped photon emission and absorption with itinerant photon–qubit interaction, we realize two distributed quantum computing primitives: an arbitrary-phase controlled-Z gate and deterministic single-shot readout of a spatially remote qubit with classical error correction.

\section{Bidirectional photon transfer}

{\it Experimental setup ---} We operate a two-node prototype quantum network based on a cryogenic microwave link \cite{Schaer2026}.
Two dilution refrigerators form the nodes of the network and are connected via a 30-meter-long cooled quantum channel housing a rectangular aluminum waveguide. The waveguide allows us to send microwave photons between the nodes in both directions, and is kept superconducting at a temperature below 60~mK along its entire length, with measured attenuation below 1~dB/km.

\begin{figure}
  \begin{center}
  \includegraphics[width=1\columnwidth]{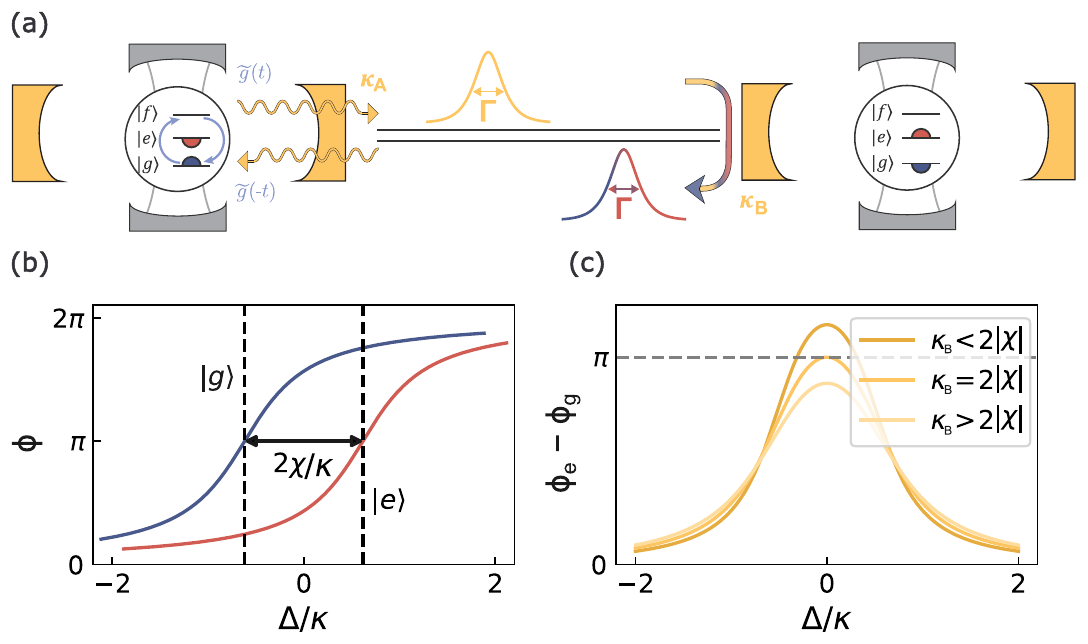}
  \caption{\textbf{Bidirectional photon transfer protocol. (a)} A schematic representation of the protocol. Node A emits a photon by inducing a tunable coupling $\tilde{g}(t)$ between the emitter qubit and the transfer resonator. The photon travels to node B, gets reflected off it and acquires a conditional phase in the process. Then node A reabsorbs the photon by applying an inverse coupling $\tilde{g}(-t)$. \textbf{(b)} Phase acquired by a communication microwave photon as a function of its detuning $\Delta$ from the remote transfer resonator frequency $\omega_{\textrm{R}_{\textrm{B}},\ket{g}}$. The two curves correspond to the remote qubit being in the ground (blue) and excited (red) state. \textbf{(c)} Conditional phase acquired by the photon. The three curves correspond to different ratios between the remote qubit's dispersive shift $\chi$ and the remote transfer resonator's decay rate $\kappa_{\rm B}$.
  }
  \label{fig:bidir_transfer}
  \end{center}
\end{figure}

The two nodes contain nominally-identical superconducting chips cooled down to 15~mK, see Fig.~\ref{fig:bidir_transfer} (a). Each chip has a frequency-tunable transmon qubit coupled to the communication waveguide via an on-chip coplanar (\textit{transfer}) resonator, a dedicated Purcell filter and a superconducting coaxial cable \cite{Storz2023}. We use the ground $\ket{g}$ and first two excited states $\ket{e},\ket{f}$ of the transmons to perform the protocols introduced below. The qubit frequencies are tuned in-situ using fast flux-bias pulses applied through a $\sigma_z$ control line, while state rotations are driven by microwave pulses applied through a charge ($\sigma_x$) control line. A second resonator-filter pair is used for dispersive readout; see Appendix~\ref{app:device} for more details.

{\it Bidirectional photon transfer ---} At the heart of our scheme lies bidirectional transfer of a shaped microwave photon between the two nodes of the network.
We begin the photon transfer protocol by performing active all-microwave reset~\cite{Magnard2018} on an emitter qubit (node A in Fig.~\ref{fig:bidir_transfer} (a)) in order to initialize it in the ground state $\ket{g}$. We then excite the transmon into the $\f$ (second excited) state \cite{Motzoi2009DRAG}, and perform the emission of a microwave photon \cite{Pechal2014} into the communication channel using a sideband transition $\ket{f0}$-$\ket{g1}$ induced by a microwave pulse. Here, $\ket{0}$,$\ket{1}$ denote the Fock states of the transfer resonator. For the photon envelope, we use the hyperbolic secant function
\eqnl{eqn:ph_shape}{f(t) = \frac{\sqrt{\Gamma}}{2}\,\textrm{sech}\left(\frac{\Gamma t}{2}\right)} with a controllable photon emission rate $\Gamma/2\pi$ up to 24~MHz, which corresponds to the photon length of as low as 120~ns. 

After emitting the photon, we allow it to propagate to the opposite node (node B) of the cryogenic quantum network, and get reflected off the transfer resonator $\mbox{Res}_B$ connected to the waveguide. After reflection, the photon propagates back to the emitter node, where we deterministically reabsorb it using a reverse $\ket{f0}$-$\ket{g1}$ transition \cite{Kurpiers2018}.

The choice of a time-symmetric photon envelope as per Eq.~\eqref{eqn:ph_shape} is motivated by the requirement to reabsorb the photon at the end of the transfer protocol, which is experimentally convenient to do by applying a time-reversed $\ket{f0}$-$\ket{g1}$ emission pulse \cite{Kurpiers2018, Hernandez2026}. We detail the photon emission and reabsorption processes in Section~\ref{sec:Variable_frequency}.

During the reflection at node B, the photon enters the transfer resonator in that device and experiences a phase response that depends on the state of the qubit, by virtue of dispersive interaction \cite{Duan2004}. Let us consider it in some detail. A photon reflecting off a resonator acquires a phase which depends on the frequency $\omega$ of the photon itself, the frequency $\omega_{\textrm{R}_{\textrm{B}}}$ of the resonator, and the system linewidth $\kappa_{\rm B}$. The phase is given by the following expression:
\begin{equation}
    \label{eqn:scatt_phi}
    e^{i\phi(\omega)}=\frac{i(\omega-\omega_{\textrm{R}_{\textrm{B}}})+\kappa_{\rm B}/2}{i(\omega-\omega_{\textrm{R}_{\textrm{B}}})-\kappa_{\rm B}/2}.
\end{equation}

As the frequency of the remote resonator $\omega_{\textrm{R}_{\textrm{B}}}$ depends on the state $\ket{\psi}_{\rm B}$ of the remote communication qubit via the dispersive shift $\chi$, the expression \ref{eqn:scatt_phi} yields two different phase responses depending on the state ($\ket{g}$ or $\ket{e}$) of the remote qubit, see~Fig.~\ref{fig:bidir_transfer} (b). We can consider the phase acquired by the microwave photon to be a sum of two contributions: an unconditional phase $\phi(\omega, \ket{g})$, which the photon acquires if the remote qubit is in the ground state, and the conditional phase 
\eqnl{eqn:cond_phase}{\Delta \phi = \phi(\omega, \ket{e}) - \phi(\omega, \ket{g}),} 

which is added if the remote qubit is in the excited state. The magnitude of the conditional phase $\Delta \phi$ can be set during the device fabrication process by choosing an appropriate dispersive shift $\chi$ and linewidth $\kappa_{\rm B}$ of the system, and tuned in the experiment by emitting communication microwave photons at different frequencies and hence adjusting the photon-resonator detuning. The maximum conditional phase $\Delta \phi$ is achieved when the communication photon is emitted at the center frequency between $\omega_{\textrm{R}_{\textrm{B}},\ket{g}}$ and $\omega_{\textrm{R}_{\textrm{B}},\ket{e}}$ and exceeds $\pi$ if the dispersive shift exceeds half the resonator linewidth, 
\eqn{|\omega_{\textrm{R}_{\textrm{B}},\ket{e}} - \omega_{\textrm{R}_{\textrm{B}},\ket{g}}| = 2|\chi| \geq \kappa_{\rm B}.}
Figure~\ref{fig:bidir_transfer} (c) depicts the conditional phase $\Delta \phi$ as a function of the photon-resonator detuning for a fixed dispersive shift $\chi$ and a variable linewidth $\kappa_{\rm B}$.


Tuning the conditional phase to $\pi$ allows us to implement a remote controlled-Z ($CZ$) gate between the two nodes of the network, and having the ability to choose the conditional phase at will enables an arbitrary-phase controlled-phase ($CPHASE$) gate. Hence it is important to be able to emit and reabsorb microwave photons at different frequencies from node A.

The bidirectional photon transfer protocol is characterized by its {\it transfer efficiency}, which we define as the population of the $\ket{f}$ state of the emitter qubit recovered after reabsorbing the photon. This definition is directly inherited from the unidirectional photon transfer between the two network nodes, for which transfer efficiencies exceeding 84.5\% were achieved in prior work \cite{Storz2024, Kulikov2024}. For bidirectional transfer, the maximum efficiency is 67\%, obtained for resonant photons. While the waveguide connecting the two nodes contributes less than 1 percentage point to the infidelity in both cases, the main contributors are the microwave components connecting the waveguide to the on-chip transfer resonators. These include coaxial-to-waveguide adapters, coaxial cables, coaxial-to-PCB (printed circuit board) connectors, PCB lines, and the wire bonds connecting the PCB lines to the chip. The quality metrics for the unidirectional and bidirectional photon transfer are consistent with each other, as in the bidirectional photon transfer protocol the photon traverses the lossy components of the network twice.

\section{Variable-frequency photon emission}
\label{sec:Variable_frequency}

A natural solution to tuning the conditional phase acquired by the communication microwave photons is being able to emit them at different frequencies. As our architecture features fixed-frequency transfer resonators, it is necessary to emit the photons detuned from the central resonator frequency \cite{Miyamura2025, Pernas2026}. More formally, we would like to emit a time-symmetric photon with constant frequency $\omega_{\textrm{R}} + \delta\omega$, and having a constant phase.
Previous theoretical treatments of the problem replace the microwave pulse used for emission by a baseband-tunable coupling between a two-level system and a resonator, hence ignoring the AC Stark shift induced by the microwave pulse \cite{Pernas2026}. Moreover, most experimental demonstrations use a microwave drive resonant with the f0-g1 transition, i.e. only producing photons resonant with the transfer resonator \cite{Kurpiers2018, Pechal2014}. Here we present an operational first-order solution enabling emitting off-resonant photons within the linewidth of the resonator.

\begin{figure}
  \begin{center}
  \includegraphics[width=1\columnwidth]{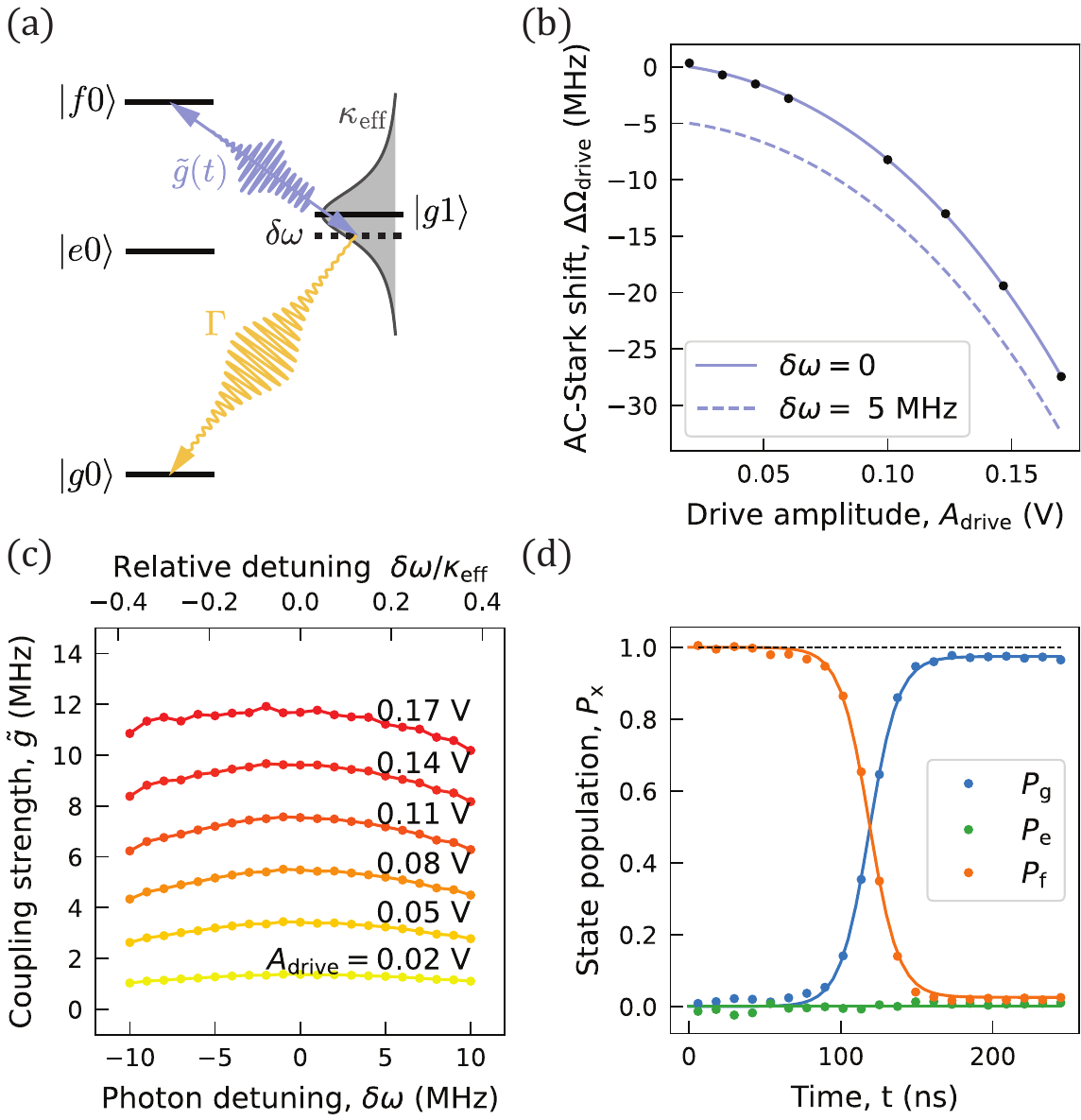}
  \caption{\textbf{Principles of photon emission. (a)} Energy level diagram of the photon emission process. The joint eigenstates of the transmon-resonator system are depicted in solid black lines. The Raman $\ket{f0}-\ket{g1}$ transition is driven with a modulated microwave pulse, creating a time-dependent coupling $\tilde{g}(t)$ (blue). The drive is detuned by $\delta\omega$ from the bare $\ket{f0}-\ket{g1}$ transition frequency (dashed line). The resonator decays into the waveguide, yielding a travelling microwave photon with a linewidth $\Gamma$ (orange). \textbf{(b)} AC Stark shift calibration result with $\Omega_{\textrm{drive}}=5.834\,\textrm{GHz}$. The markers correspond to the calibration data, the solid line is a fit with an even polynomial, and the dashed line shows the same fit shifted by $\delta\omega = 5$~MHz. \textbf{(c)} Induced $\ket{f0}-\ket{g1}$ coupling strength as a function of $\delta\omega$ for various drive amplitudes.
  \textbf{(d)} Emitter qubit state population during the shaped photon emission process, for $\delta\omega = 0 \,\textrm{MHz}$.
  }
  \label{fig:f0g1_scheme}
  \end{center}
\end{figure}

A key instrument allowing us to emit a microwave photon with shaped temporal profile is the ability to induce tunable coupling between the communication qubit and the transmission line. We apply a microwave signal modulated in amplitude and phase and resonant (up to a small detuning $\delta\omega$) with the $\ket{f0}$-$\ket{g1}$ transition of the joint transmon-resonator system (see Fig.~\ref{fig:f0g1_scheme} (a)) at the frequency of 
$\Omega_{\textrm{drive}}^0 -\delta \omega=\omega_{ge} + \omega_{ef} - \omega_{\textrm{R}}-\delta \omega$.
The term $\delta\omega$ introduces a detuning from the unperturbed $\ket{f0}$-$\ket{g1}$ transition. To first order, this allows us to emit a photon with carrier frequency $\omega_{\textrm{R}} + \delta\omega$, as expected from energy conservation.
In the rotating frame of the drive ($\delta \omega = 0$), the states $\ket{f0}$ and $\ket{g1}$ are resonant and are coupled via a second-order effective coupling, which is linear in the amplitude of the drive to first order \cite{Pechal2014}.
The combination of the tunable coupling with the spontaneous decay from the effective resonator-Purcell system to the transmission line ($\ket{g1} \rightarrow \ket{g0}$) at the independently-measured decay rate $\kappa_{\rm A}$ allows us to engineer a desired photon envelope.


Emitting a photon at constant frequency and phase is important to preserve the symmetry of its envelope function and reach a high absorption fidelity. While nominally driving the $\ket{f0}$-$\ket{g1}$ transition at the frequency of $\Omega_{\textrm{drive}}^0$ with $\delta \omega = 0$ would allow emitting a photon with the frequency of $\omega_{\rm R_A}$, this transition exhibits an AC Stark shift depending on the strength of the microwave drive. Hence the complete expression for the drive frequency reads
\eqn{\Omega_{\textrm{drive}} = \Omega_{\textrm{drive}}^0 + \Delta\Omega_{\textrm{drive}} - \delta\omega}

We employ a three-step calibration process to take these AC Stark shifts into account. The $\ket{f0}$-$\ket{g1}$ transition and shaped photon emission calibration is discussed in more detail in Appendix~\ref{app:f0g1_calibration} and in previous publications \cite{Pechal2014,Magnard2018,Kurpiers2018}.
The calibration procedure is as follows:

\begin{enumerate}
    \item 
    AC Stark shift: We calibrate and establish a bijection between the $\ket{f0}$-$\ket{g1}$ transition frequency and the amplitude of the corresponding microwave drive $\Delta\Omega_{\textrm{drive}}\leftrightarrow A_{\textrm{drive}}$, solid line in Fig.~\ref{fig:f0g1_scheme} (b).
    
\end{enumerate}

\begin{tcolorbox}[mybox, title={for each $\delta\omega$}]
\begin{enumerate}[resume]
    \setcounter{enumi}{1}
    \item 
    Detuning $\delta\omega$: We set the drive frequency to be $\Omega_{\textrm{drive}}^0 + \Delta\Omega_{\textrm{drive}} - \delta\omega$ for all amplitudes $A_{\textrm{drive}}$, dashed line in Fig.~\ref{fig:f0g1_scheme} (b).
    \item
    Rabi rate: We calibrate and establish a bijection between the amplitude of the detuned $\ket{f0}$-$\ket{g1}$ microwave drive and the induced coupling rate $A_{\textrm{drive}}\leftrightarrow \tilde{g}$. Results for all $\delta\omega$ are shown in Fig.~\ref{fig:f0g1_scheme} (c).
    \item
    Verification: We verify the generation of a shaped photon by truncating the emission pulse at a variable time and reading out the source qubit immediately afterwards, comparing the population of the transmon states to those predicted by theory for the target photon envelope $f(t)$ \cite{Hernandez2026}.
    We typically reach an emission fidelity above 98\%. An example measurement is shown in Fig~\ref{fig:f0g1_scheme} (d).
\end{enumerate}

\end{tcolorbox}

The frequency shift $\delta\omega$ changes the induced coupling between the $\ket{f0}$ and $\ket{g1}$ levels significantly and differently for each drive amplitude, as the driving is off-resonant.

The target photon envelope determines the required coupling $\tilde{g}(t)$. Using the calibrated relation $A_{\textrm{drive}}\leftrightarrow \tilde{g}$, we are then able to determine the required drive amplitude waveform $A_{\textrm{drive}}(t)$; The AC Stark calibration allows assigning the corresponding instantaneous drive frequency. In this way, we construct a variable-frequency and variable-amplitude microwave pulse emitting a photon at a frequency $\omega_{\textrm{R}}+\delta\omega$. It is important to note that the reason why the $\ket{f0}$-$\ket{g1}$ microwave drive has to have a frequency chirp (frequency varied in time) is the presence of the AC Stark shifts. Without the AC Stark shifts, applying a constant-frequency $\ket{f0}$-$\ket{g1}$ drive would have resulted in a constant-frequency photon.
The photon shaping method described above with $\delta \omega = 0$ has been used in several quantum communication experiments on the circuit QED platform~\cite{Kurpiers2018,Storz2023,Ilves2020} for the purpose of entanglement distribution between remote superconducting nodes. 


\begin{figure}
  \begin{center}
  \includegraphics[width=1\columnwidth]{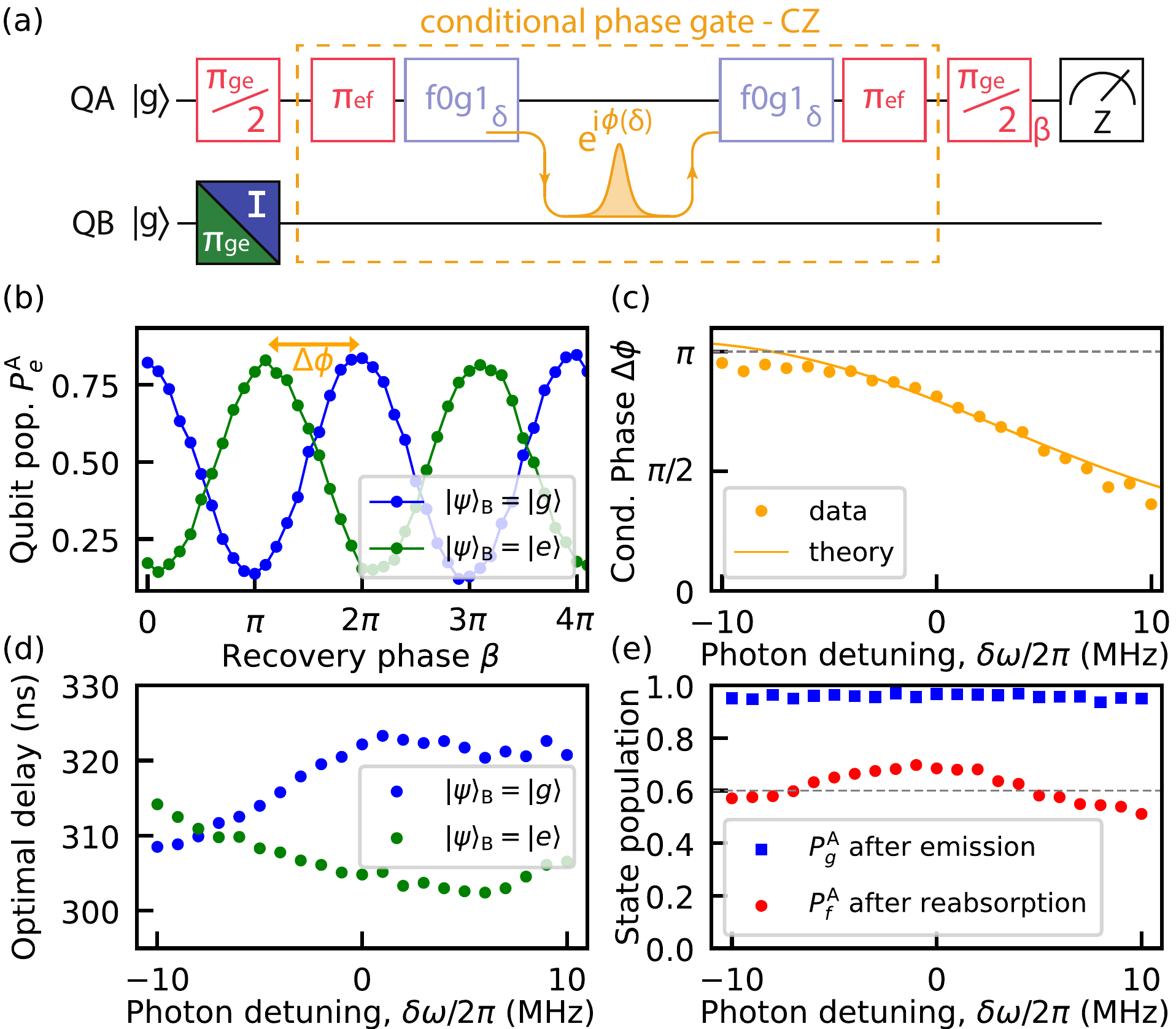}
  \caption{\textbf{Conditional phase tunability.}
  \textbf{(a)} Gate sequence for the characterization of the conditional phase. Single-qutrit gates in transmon A in red, $f0{\rm -}g1$ emission and absorption operations in light blue, photon in yellow, readout operation in black, single-qubit gates in transmon B in dark blue/green.
  \textbf{(b)} Two-curve method for measuring the conditional phase $\Delta \phi$ of the communication photon; see text for details.
  \textbf{(c)} Conditional phase $\Delta \phi$ acquired by the communication photon as a function of the photon detuning $\delta\omega$ from the resonator frequency $\omega_{\textrm{R}}$. Markers are measured data; solid line is the theoretical prediction from the $\delta\omega = 0$ point and equation~\eqref{eqn:scatt_phi}.
  \textbf{(d)} Optimal reabsorption delay of the photon as a function of $\delta\omega$ for the remote qubit being in $\ket{g}$ (blue) or $\ket{e}$ (green) states.
  \textbf{(e)} State population of the $\ket{g}$ state ($\ket{f}$ state) of the emitter qubit after the photon emission (reabsorption) as a function of $\delta\omega$.
  }
  \label{fig:phase_tunability}
  \end{center}
\end{figure}

The described procedure is a practical first-order solution to the full Hamiltonian problem. It allows emitting a microwave photon detuned from the emitter's transfer resonator in a narrow range of approximately $\kappa_{\rm A}$, from $-\kappa_{\rm A}/2$ to $\kappa_{\rm A}/2$. A recent work \cite{Miyamura2025} achieves a similar goal through a different strategy -- characterizing the generated photon amplitude and frequency from the drive-tone parameters, and using this characterization to emit photons with the desired spectral and temporal profiles. In our case, the photon is emitted straight into the bidirectional communication line, which precludes {\it in situ} measurement of its shape or spectrum. We therefore confirm the success of frequency tuning by measuring the tunability range of the conditional phase acquired by the photon in a bidirectional photon transfer experiment using the remote node.


\begin{figure*}
  \begin{center}
  \includegraphics[width = \linewidth]{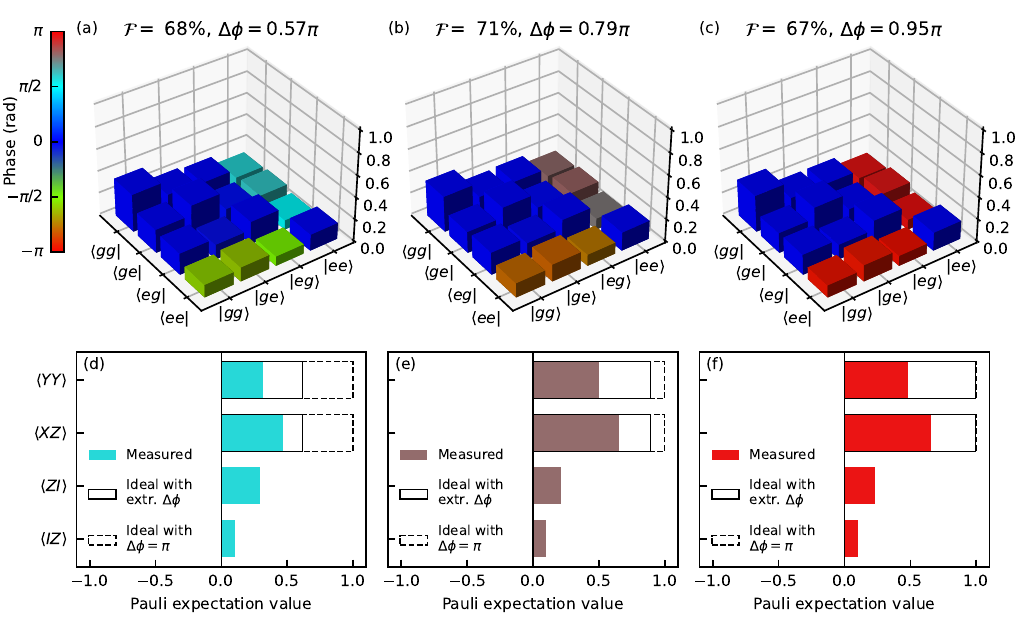}
  \caption{\textbf{Arbitrary-phase controlled-Z gate at different photon detunings.} \textbf{(a)--(c)} Reconstructed two-qubit density matrices after applying the remote controlled-phase gate to the initial state $\ket{++}$ for photon detunings of $\delta\omega  = - 5$~MHz \textbf{(a)}, $\delta\omega  = 0$ \textbf{(b)} and $\delta\omega  = 5$~MHz \textbf{(c)}. The extracted state fidelities are $\mathcal{F}=68\%$, $71\%$, and $67\%$, and the corresponding conditional phases are $\Delta\phi=0.57\pi$, $0.79\pi$, and $0.95\pi$. The change in the phase of the off-diagonal elements of the density matrices demonstrates tunability of the remote controlled-phase operation. \textbf{(d)--(f)} Selected Pauli expectation values extracted from the same tomographic data. Measured values are compared with the ideal output state for the experimentally extracted conditional phase and with the ideal output state for $\Delta\phi=\pi$. The observed single-qubit $Z$ components are consistent with photon loss predominantly affecting the emitter qubit during the bidirectional photon-transfer protocol.}
  \label{fig:CZ_DMs}
  \end{center}
\end{figure*}

{\it Tuning the conditional phase ---} We employ a Ramsey-like sequence to determine the conditional phase $\Delta \phi$ (see eq.~\eqref{eqn:cond_phase}) of the communication photon. We prepare a $\ket{+}$ state in the emitter qubit \cite{Magnard2018, Motzoi2009DRAG}, A, and perform a bidirectional photon transfer protocol with detuning $\delta\omega$ in the $f0-{\rm}g1$ pulse, followed by a $(\pi/2)$ pulse with variable phase $\beta$, see Fig.~\ref{fig:phase_tunability} (a). The subsequent readout of the emitter qubit measures the population of the $\ket{e}$ state, which oscillates as a function of the recovery phase $\beta$, see blue curve in Fig.~\ref{fig:phase_tunability} (b), for $\delta\omega/2\pi=-3~{\rm MHz}$.
The experiment is repeated applying a $\pi$-pulse to excite qubit B to its $\ket{e}$ state before the bidirectional photon transfer, resulting in oscillations with a different phase (green). The phase difference between the two curves is the conditional phase $\Delta \phi$.

We tune the photon across a 20~MHz range centered at the frequency of the transfer resonator $\omega_{\textrm{R}}$ of the local node using the method outlined above, and measure the conditional phase $\Delta \phi$ acquired by the photon at each detuning. The measured phase approaches $\Delta\phi = \pi$ around $\delta\omega = -8~{\rm MHz}$, and decreases smoothly to below $\pi/2$ at positive detunings, see Fig.~\ref{fig:phase_tunability} (c). These results are in good agreement with the theoretical prediction obtained from equation \eqref{eqn:scatt_phi}, and consistent with the qualitative dependence illustrated in Fig.~\ref{fig:bidir_transfer} (c). We attribute the slight deviations at higher detunings to the approximate nature of the variable-frequency photon emission method. Nevertheless, we successfully tune the conditional phase in a wide range covering an interval $[\pi/2,0.95\pi]$.

Another effect allows us to indirectly verify the frequency tuning of the communication photon in an independent manner. We calibrate the optimal reabsorption time, i.e. the one maximizing the absorption efficiency, when the remote qubit is in the state $\ket{g}$, and observe a strong dependence on the photon detuning $\delta\omega$, see blue markers in Fig.~\ref{fig:phase_tunability} (d). Namely, the optimal reabsorption time in this case is maximal at the detuning of $\delta\omega=$1.5~MHz, which corresponds to the communication photon being resonant with the remote transfer resonator, and decreases towards negative detunings. While the return travel time of the wavepacket in the communication channel changes negligibly within the narrow frequency range of 20~MHz realized here, the interaction time with the remote transfer resonator, or the group delay $\tau_{\rm g} = d\phi/d\omega$, changes significantly. This is due to the phase response $\phi(\omega)$ of the resonator, acquired by the photon upon reflection, which features a nonlinear frequency dependency, see Eq.~(\ref{eqn:scatt_phi}). We also characterize the optimal delay when the remote qubit is in state $\ket{e}$, see green markers in Fig.~\ref{fig:phase_tunability} (d). The optimal delay times for the two remote states, $\ket{g}$ and $\ket{e}$, become equal at the detuning of -8~MHz, which corresponds to the point where the frequency of the communication photon is exactly in the middle between $\omega_{\textrm{R}_{\textrm{B}},\ket{g}}$ and $\omega_{\textrm{R}_{\textrm{B}},\ket{e}}$. Note how the maximal conditional phase in Fig.~\ref{fig:phase_tunability} (c) is achieved with the same photon frequency, in agreement with eq.~\eqref{eqn:scatt_phi} and Fig.~\ref{fig:bidir_transfer} (c).

Finally, we characterize the bidirectional photon transfer efficiency by measuring the population of qubit A after the absorption operation, for a variable photon frequency, see Fig.~\ref{fig:phase_tunability} (e). With a maximum value of 64\% achieved with the resonant photon, the efficiency stays above 57\% for the entirety of the tuning range (the mechanisms influencing this efficiency are discussed in App.~\ref{app:errors}), which together with the conditional phase tunability enables the implementation of an arbitrary-phase CZ gate.

\section{Distributed quantum computing primitives}

{\it Conditional phase gate ---} We can now demonstrate a native implementation of a remote entangling gate between the nodes of our network. We use the conditional phase acquired by the communication microwave photon to realize a controlled-phase (CPHASE) gate, whose angle can be tuned arbitrarily in the range $[\pi/2,0.95\pi]$.

\begin{figure*} 
  \begin{center} \includegraphics[width = \linewidth]{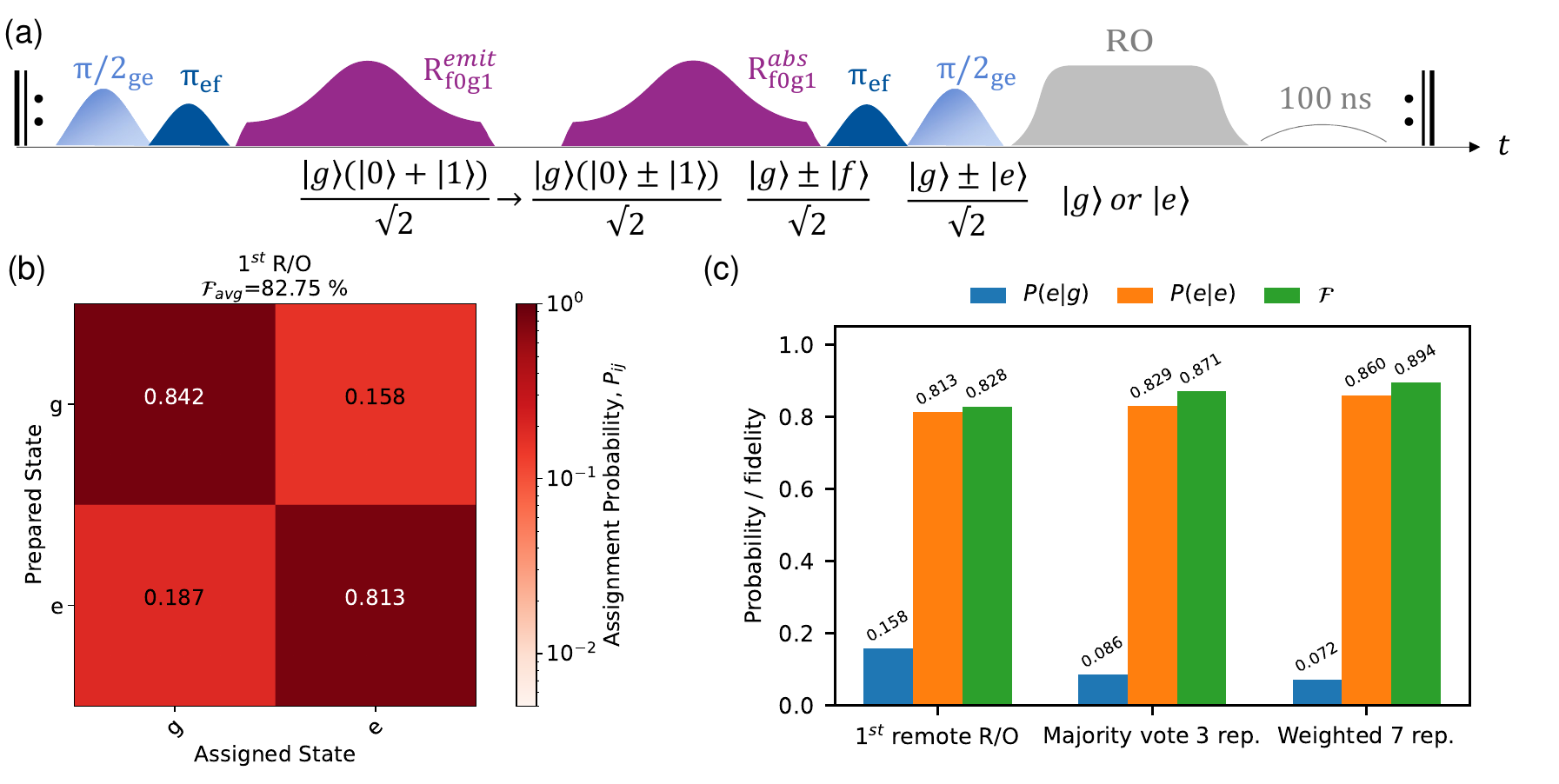}
  \caption{\textbf{Deterministic single-shot readout of a quantum bit located 30~m away. (a)} Remote readout protocol. Light blue corresponds to the qubit $\ket{g}$-$\ket{e}$ rotations, dark blue -- to the qubit $\ket{e}$-$\ket{f}$ rotations, and purple -- to $\ket{f0}$-$\ket{g1}$ transition microwave drive. States under the line represent the joint emitter-photon state under ideal conditions. \textbf{(b)} State assignment matrix for a single run of a remote readout protocol. \textbf{(c)} Classical error correction of the remote readout. Blue and orange bars correspond to the probability to identify the remote state as $\ket{e}$ while the prepared state is $\ket{g}$ or $\ket{e}$, respectively. Green bars correspond to the measurement fidelity. The first block reflects the performance of a single-shot remote readout protocol run once. The second block shows the performance improvement if the protocol is repeated three times in a row with majority voting. The last block shows the best achieved performance with repeated remote readout done with seven repetitions and optimal weights.}
  \label{fig:Remote_readout}
  \end{center}
\end{figure*}

We demonstrate the phase tuning directly by preparing the two qubits in a separable superposition state $\ket{++}$, and then performing the CPHASE gate at various photon detunings. We characterize the entanglement and conditional phase after applying the gate by analyzing the density matrix $\rho$ of the system, which we measure using two-qubit tomography. Three examples of the resulting states for different photon frequencies are shown in Fig.~\ref{fig:CZ_DMs}.

For all three examples the conditional phase, extracted from the phase of $\bra{gg}\rho\ket{ee}$, coincides with the conditional phase extracted via the Ramsey-like sequence and shown in Fig.~\ref{fig:phase_tunability} (c). We calculate the fidelity of the resulting state relative to an ideal state with the corresponding phase. The fidelity of the two-qubit state peaks at $\mathcal{F} = 71\%$ when the photon is emitted resonantly, and decreases slightly to $\mathcal{F} = 67\%$ towards the ends of the tunability range. The CPHASE gate fidelity is mainly limited by the bidirectional photon transfer efficiency (see Fig.~\ref{fig:phase_tunability} (e)), and therefore by the photon loss in the communication channel, see App.~\ref{app:errors} for a more detailed breakdown of the errors in this operation.

The native implementation of remote arbitrary-phase CZ gate allows unifying the nodes of a quantum network in a singular distributed quantum processor, and executing algorithms in the same manner as on a single chip \cite{Barenco1995}. Our results embody the first demonstration of a remote two-qubit gate for the circuit QED platform based on the exchange of itinerant photons, and offer a longer-range alternative to recently-implemented remote two-qubit gates mediated by cable-resonator modes \cite{Niu2023, Heya2025}. Direct all-microwave protocols may be useful for local networks of superconducting qubits as an alternative to transduction to the optical domain, where achieving high-efficiency conversion remains challenging~\cite{Mirhosseini2020, Zhao2025b}, or teleportation-based routines \cite{Qiu2025a}. Finally, even though a fixed-phase controlled-Z gate together with single-qubit operations would already generate a complete gate set for distributed quantum computing \cite{Barrett2005}, having the ability to tune the phase can greatly decrease the amount of operations in certain cases when performing distributed algorithms, as any two-qubit unitary can be efficiently implemented with arbitrary-phase CZ gate \cite{Lacroix2020}.

{\it Remote readout ---} The second distributed quantum computing primitive we implement on the basis of the bidirectional photon transfer is the deterministic single-shot readout of a remote quantum bit. We excite the emitter qubit in the superposition state $\ket{+}$, and then perform a CZ gate between the local and remote qubits. Setting the conditional phase of the arbitrary-phase gate to $\pi$ allows us to consider the gate application as flipping the sign of the superposition to $\ket{-}$ in the case where the remote qubit is in the excited state. Hence a final $(\pi/2)_y$ pulse yields $\ket{g}$ in the case where the remote qubit is in $\ket{g}$, and vice versa. The full remote readout protocol is shown in Fig.~\ref{fig:Remote_readout} (a). Note that during the protocol we fully disentangle the emitter qubit from the itinerant radiation mode. Hence in case of photon loss the emitter qubit does not leave the computational subspace, but rather ends up in a mixed state before the measurement.

We experimentally implement the protocol and obtain a single-shot state assignment matrix, see Fig.~\ref{fig:Remote_readout} (b). The remote readout fidelity for a single protocol execution reaches $83\%$, with the dominant source of errors being photon loss in the channel, see App.~\ref{app:errors}. We can, however, address this dominant error source by the means of classical error correction. 

The pulse that probes the qubit state consists of a single microwave photon, leading to remote readout being highly-QND (quantum nondemolition), with no possibility to excite any higher transmon levels \cite{Sank2016, Chapple2025a}, like during a conventional local dispersive readout \cite{Swiadek2024}. Hence the repeatability of the remote readout protocol is only limited by the lifetime of the remote qubit, with the successive executions yielding identical results subject to noise and loss sources. Repeating the protocol three times and doing majority voting on the results already allows defining the state of the remote qubit correctly unless a communication photon was lost at least twice out of three repetitions, i.e. limiting the errors to the second order in the error probability.

In order to implement the repeated remote readout, one extension to the single run of the protocol has to be defined: we allow the local qubit to start not only in $|g\rangle$, but also in $|e\rangle$. It is straightforward to see that in this case the phases of all superposition state during the execution of the protocol are inverted, and therefore we infer the state of the remote qubit to be the opposite from the local readout.
Hence, in the full protocol, the inference on the state of the remote qubit is the sum of the starting state and the local readout modulo 2, taking $|g\rangle = 0$ and $|e\rangle = 1$.

Finally, we can perform more iterations of the remote readout protocol, calculate optimal weights and do thresholding in a similar fashion to a local readout operation. The comparison between three fidelities: that of a single remote readout, of the three-times-repeated remote readout with majority voting, and that of the 7-times repeated protocol with optimal weights is shown on Fig.~\ref{fig:Remote_readout} (c). The classical error correction halves the remote readout error, and is projected to have even better performance improvement as the errors of a single protocol run decrease with lower photon losses.


\section{Software integration and applications}

\begin{figure}
  \begin{center}
  \includegraphics[]{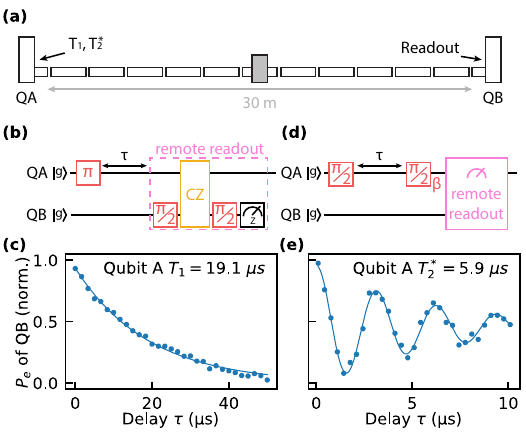}
  \caption{\textbf{Seamless integration of the remote readout operation, for the remote characterization of qubit coherence times.}
  \textbf{(a)}~Setup schematic. The dispersive readout operation is executed in a transmon 30~m away from the qubit whose coherence times are to be measured.
  \textbf{(b)}~Gate sequence for remote $T_1$ measurement. Single-qubit gates in red, remote CZ gate from Fig.~\ref{fig:phase_tunability} (a) in yellow, readout in white. The delay $\tau$ indicates the separation between the first two single-qubit gates.
  \textbf{(c)}~Population of qubit A measured remotely with sequence (b) vs. the delay $\tau$, normalized to a calibration measurement with $\tau=0$. Markers are measured data, solid line is a fit to an exponential function, used to infer the lifetime $T_1=19.1~\upmu{\rm s}$ of qubit A.
  \textbf{(d)}~Gate sequence for remote $T_2$ measurement. Single-qubit gates in red, remote readout operation from (b) in pink. The delay $\tau$ indicates the separation between the first two single-qubit gates.
  \textbf{(e)}~Population of qubit A measured remotely after the Ramsey sequence in (d) vs. the delay $\tau$, normalized to a calibration measurement with $\tau=0$. Markers are measured data, solid line is a fit to an exponentially damped oscillation, used to infer the dephasing time $T_2=5.9~\upmu{\rm s}$ of qubit A.
  }
  \label{fig:seamless_integration}
  \end{center}
\end{figure}

{\it Seamless integration ---} As our quantum network features stable phase coherence across the two nodes, we can perform the calibration and characterization of the remote readout and proceed to integrate it into the existing measurement software as a standard operation alongside with local readout or single-qubit gates. This enables running a quantum algorithm on a networking node, while performing all readout operations remotely via a different node. We demonstrate the functionality of such seamless integration by executing two calibration quantum experiments, $T_1$ decay and Ramsey oscillations. In both experiments, we read out the qubit in the node A, where the experiments are performed, remotely via the system in the node B, located 30 meters away, see Fig.~\ref{fig:seamless_integration} (a). 

For the $T_1$ measurement, we excite qubit A, wait for a variable delay $\tau$ and then execute the remote readout operation from qubit B, see Fig.~\ref{fig:seamless_integration} (b). We normalize the extracted population using two additional reference measurements, where qubit A is initialized to the states $\ket{g}$ or $\ket{e}$ and immediately read out remotely via B. We obtain an exponentially decaying excited-state population in qubit B, which we fit and extract the lifetime $T_1$ of qubit A, see Fig~\ref{fig:seamless_integration} (c). Analogously for the $T_2$ measurement, we realize a Ramsey sequence in qubit A with a recovery phase $\beta\neq0$, and whose readout is executed remotely from qubit B, see Fig.~\ref{fig:seamless_integration} (d). We obtain the damped oscillation in Fig~\ref{fig:seamless_integration} (e), normalized with the same method, from which we extract the dephasing time $T_2$ of qubit A.

Both demonstrated networking primitives require active control only on the emitter node, while the remote node remains fully passive. It simplifies the network engineering as no precise temporal synchronization between the nodes is required in order to successfully execute a distributed algorithm. Indeed, only the sequence of operations (qubit gates, readout, reset) has to be enforced, but not the exact timings within few ns or less. It could be particularly useful in larger networks, where clock synchronization becomes an issue \cite{Valivarthi2022}.

{\it Quantum hacking ---} An interesting feature of the presented remote readout protocol we cannot omit mentioning stems from the fact that the protocol is fully passive on the node whose qubit is being read out. The active node can perform the remote readout protocol at will, with no requirement to synchronize any control pulses or gates between the two nodes, and even to ask permission from the remote node. While this leads to clear advantages in the distributed quantum computing case outlined previously, it opens security vulnerabilities in the network case. For instance, should an actual quantum processor have several of its quantum bits connected to an untrusted network, these quantum bits could in principle be read out from the outside at any time through a network attack. Such an attack would be realized with remote readout performed with a microwave communication photon injected into the network by an adversary.

These considerations are particularly relevant for various device-independent (DI) algorithms, such as recently-implemented device-independent quantum key distribution (DI-QKD) routines \cite{Zhang2022l, Nadlinger2022}. In such routines, if the qubits are spatially separated in different nodes of a network, a critical requirement for the security of the routine is the assumption of each node being a closed system, i.e. that no quantum or classical information could leak from it to an adversary~\cite{Acin2007b,Murta2019}. Hence, if a qubit participating in DI-QKD is connected to a cavity-interfaced bidirectional network, a remote-readout-based attack can compromise the security of the secret key generated by the DI-QKD routine. Such attacks targeting and invalidating critical assumptions underlying communication security are not new, and have been explored in the optical domain \cite{Lydersen2010,Huang2019c}. 

One potential straightforward solution in the case of superconducting architecture would be to split the purpose of quantum bits in a network-connected quantum processor into 'communication' qubits, whose sole purpose is to function as gates or ports (like Ethernet port in classical communication and networking) to send and receive quantum information, and 'computation' qubits, which will not be directly coupled to the network. Such separation would alleviate the vulnerabilities based on this and other similar protocols.

\section{Conclusions and outlook}

In this work we have realized a CZ gate between two superconducting qubits separated by 30 meters, using the round trip of a microwave photon to mediate the interaction. The conditional phase can be controlled by tuning the frequency of the emitted photon, which we have addressed with an operational calibration method. While there is concurrent work in this direction \cite{Miyamura2025}, further theoretical and experimental investigation will be required to improve the gate fidelity and simplify its calibration.

Our gate leverages the direct exchange of quantum states, requiring no frequency-tunable on-chip components, no interaction with additional memory qubits, no classical communication, no tight synchronization and no real-time classical logic. It is also deterministic and can be used between quantum nodes separated by tens of meters \cite{Storz2023, Schaer2026}. Additionally, this gate could be executed between multiple pairs of qubits in parallel using a single communication channel and frequency or orthogonal-mode multiplexing \cite{Hernandez2026}. Both operations are therefore suitable for the data-center-scale distance regime, which is not directly addressed by chip-scale or package-scale couplers. This regime is becoming increasingly relevant as industrial and academic efforts toward larger superconducting processors explore multi-cryostat and networked architectures~\cite{Heya2025,AghaeeRad2025}, aiming to integrate networking with quantum error correction and fault-tolerant protocols~\cite{Xu2022e,Ramette2023}.


Avenues for potential improvement include reducing the loss in the channel \cite{Niu2023} and gaining better control over the photon shape and frequency.
These will greatly enhance the viability of the presented protocols for data-center-scale quantum networks, where approaches based on transduction and optical communication may be too complex or unavailable in the short term. Meanwhile, protocols based on gate teleportation, frequency conversion, classical communication and probabilistic schemes like heralding, distillation or quantum repeaters will likely still be necessary for longer distances.

\newpage

\bibliographystyle{apsrev4-1}

\bibliography{RefDB/basic_refs,RefDB/extra_refs,RefDB/DIRA_refs,RefDB/qudevRefDB_part}

\newpage

\

\newpage

\section*{Acknowledgments}
The authors would like to thank Juan Jos\'e Garc\'ia-Ripoll and Guillermo F. Pe\~{n}as for helpful discussions, and Jean-Claude Besse and Andreas Wallraff for feedback on the manuscript and the figures. 
The work was supported through resources supplied by the Quantum Device Lab at the Department of Physics of ETH Zurich.

\section*{Competing Interests}
The authors declare no competing interests.

\section*{Data Availability}
All data is available from the corresponding authors upon reasonable request.


\section {Supplemental information}

\setcounter{secnumdepth}{2}

\appendix
\section{Device Parameters}
\label{app:device}

\begin{figure}[hbtp]
    \centering
    \includegraphics[width = \linewidth]{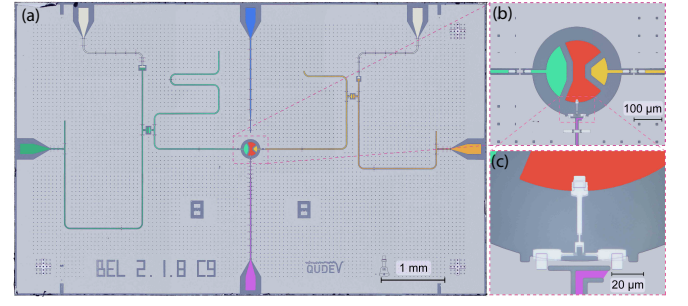}
    \caption{(a) False-colored micrograph of the superconducting chip used in the experiment, taken from \cite{Storz2023}. The transmon-style qubit in the center (red) is coupled to the readout circuitry (green), transfer circuitry (orange), flux control line (purple) and microwave drive line (blue). The gray ports (top left and right) are the inputs of the readout and transfer circuitry. (b) Enlarged view of the qubit with the capacitive coupling pads to the readout and transfer circuitry. (c) Enlarged view of the SQUID loop of the qubit.}
    \label{fig:chip_layout}
\end{figure}

The superconducting chips used in this work have dimensions of 4.3~mm $\times$ 7~mm and are fabricated on silicon substrates patterned with a 150~nm-thick niobium film. The coplanar waveguide structures and transmon geometry, shown in Fig.~\ref{fig:chip_layout} (a,b), are defined using photolithography and reactive-ion etching. The airbridges are added in a two-layer resist photolithography process. Finally, the Dolan bridge Josephson junctions, shown in Fig. \ref{fig:chip_layout} (c), are realized using electron-beam lithography followed by double-angle shadow evaporation.

Each chip is mounted on a copper printed circuit board, wire-bonded, and installed in a shielded sample holder at the base stage of a dilution refrigerator. The readout and control ports are connected to room-temperature electronics located at the corresponding network node. The transfer input port is terminated by an impedance-matched load thermalized to the base temperature of the refrigerator, while the transfer output is connected to the cryogenic quantum link through an SMA-to-SMP adapter and a niobium–titanium coaxial cable.

The experiment employs two nominally identical devices. The parameters of the readout and transfer circuits are extracted from transmission spectroscopy measurements with the qubit prepared in the $\ket{g}$ and $\ket{e}$ states, following the methods of Ref.~\cite{Walter2017}. The qubit transition frequencies, coherence times, and thermal populations are characterized using standard spectroscopy and Ramsey-type measurements, as well as the method described in Ref.~\cite{Devoret2013Reset}. The measured device parameters are summarized in Table~\ref{tab:ChipParameters}. The transition frequencies, coherence times, thermal populations, and transfer-circuit parameters are measured with the qubits biased at their upper flux symmetry points. The parameters for the readout circuitry are taken at the qubits' designated readout operation point optimized for high\nobreakdash-fidelity readout, i.e. at flux-biased frequency of about $\omega_{ge}\approx$~7.2~GHz.

To realize a bidirectional quantum link suitable for remote gate and readout operations, we removed the circulator that was present in the earlier network implementation of Ref.~\cite{Storz2023}. In that configuration, the circulator enabled measurements of the temporal mode functions of emitted microwave photons and transmission spectroscopy of the transfer circuitry. Since these measurements are no longer accessible in the bidirectional configuration, the parameters marked with (*) in Table~\ref{tab:ChipParameters} are taken from Ref.~\cite{Storz2023}.

The equilibrium thermal population of the qubits is approximately 1\%, substantially lower than the value of approximately 7\% reported in Ref.~\cite{Storz2023}. We attribute this improvement to the installation of radio-frequency absorbing polyurethane foam inside the light-tight and magnetically shielded sample enclosure.

\section{Conditional phase calibration}
To measure the phase that the photon acquires when it reflects off node~B, we perform a Ramsey-like experiment. We first apply a $(\pi/2)$ pulse at the $|g\rangle \rightarrow |e\rangle$ transition frequency bringing the state of the qubit to the equatorial plane of the Bloch sphere. We then perform photon emission and reabsorption according to the protocol described in Fig.~\ref{fig:phase_tunability}. The final resonant $(\pi/2)$ pulse is applied with variable phase, corresponding to rotations about different equatorial axes of the Bloch sphere. For a fixed qubit state, changing the rotation axis changes the final qubit population as a function of the phase of the final pulse. Repeating the measurement with node~B prepared in $|g\rangle$ and $|e\rangle$ produces two oscillation curves with different phase offsets. While the individual phase offsets depend on the phase of the qubit state relative to the phase reference provided by the control electronics, the difference between them gives the conditional phase acquired during the emission–reflection–reabsorption process.

To extract the conditional phase we first estimate the qubit population for both states, $|g\rangle$ and $|e\rangle$, of the remote node and for each projection phase via repeated single-shot measurements. This gives two population curves as a function of the projection phase. We subtract the background offset and normalize the curves by the largest population difference in order to preserve only the phase information. We then take the discrete Fourier transform of each curve and identify the dominant Fourier component, corresponding to the scanned population oscillation. The phases of this Fourier component, $\phi_g$ and $\phi_e$, define the phase offsets of the two population oscillations. The conditional phase is then obtained as:

\begin{equation}
    \Delta \phi = (\phi_e - \phi_g)\  \mathrm{mod}\  2\pi
\end{equation}

To study how the conditional phase depends on the photon frequency, we perform the measurement for photon frequencies within a $\pm 12$~MHz range around the node~A resonator frequency. We observe $\Delta \phi$ changing from $\approx 0.3 \pi$ to $\approx 0.95\pi$, which is in good agreement with the conditional phases extracted from full state tomography of qubit A.

\begin{table}[t]
{\footnotesize{}}%
\footnotesize
\begin{tabular}{|ll|r|r|}
\hline
{quantity, symbol } & {unit } & {\hfill{}Node A\hfill{} } & {\hfill{}Node B\hfill{}}\tabularnewline
\hline
\hline

{qubit $ge$ transition frequency, $\omega_{ge}/2\pi$ } & {GHz } & {7.845 } & {7.901}\tabularnewline
{transmon anharmonicity, $\alpha/2\pi$ } & {MHz } & {-311 } & {-308 }\tabularnewline
{e to g level energy relaxation time, $T_{{\rm 1ge}}$ } & {$\mathrm{\mu s}$ } & {19.8 } & {25.2 }\tabularnewline
{f to e level energy relaxation time, $T_{{\rm 1ef}}$ } & {$\mathrm{\mu s}$ } & {7.6 } & {5.0 }\tabularnewline
{coherence time on $ge$ subspace , $T_{{\rm 2ge}}^{*}$ } & {$\mathrm{\mu s}$ } & {6.0 } & {6.2 }\tabularnewline
{coherence time on $ef$ subspace, $T_{{\rm 2ef}}^{*}$ } & {$\mathrm{\mu s}$ } & {4.8 } & {4.1 }\tabularnewline
{thermal excitation at equilibrium, $n_{{\rm th}}$ } & {\% } & {1.0 } & {1.5 }\tabularnewline 
{$\ket{f,0}\leftrightarrow\ket{g,1}$ transition frequency,
$\nu_{\mathrm{f0g1}}$ } & {GHz } & {5.861 } & {5.972 }\tabularnewline
\hline
{readout resonator frequency, $\omega_{r}/2\pi$ } & {GHz } & {5.632 } & {5.636 }\tabularnewline
{readout Purcell filter frequency, $\omega_{{\rm Pr}}/2\pi$ } & {GHz } & {5.620 } & {5.611 }\tabularnewline
{readout resonator/qutrit coupling, $g_{r}/2\pi$ } & {MHz } & {198 } & {224 }\tabularnewline
{readout resonator dispersive shift, $\chi_{r}/2\pi$ } & {MHz } & {-6.4 } & {-8.0 }\tabularnewline
{readout resonator/filter coupling, $J_{r}/2\pi$ } & {MHz } & {29.9 } & {26.2 }\tabularnewline
{readout Purcell filter bandwidth, $\kappa_{{\rm Pr}}/2\pi$ } & {MHz } & {95.4 } & {92.1 }\tabularnewline
{readout resonator eff. bandwidth, $\kappa_{r}/2\pi$ } & {MHz } & {35.2 } & {29.8 }\tabularnewline
\hline
{*transfer resonator frequency, $\omega_{t}/2\pi$ } & {GHz } & {9.490 } & {9.491 }\tabularnewline
{*transfer Purcell filter frequency, $\omega_{{\rm Pt}}/2\pi$ } & {GHz } & {9.489 } & {9.477 }\tabularnewline
{*transfer resonator/qutrit coupling, $g_{t}/2\pi$ } & {MHz } & {255 } & {324 }\tabularnewline
{transfer resonator dispersive shift, $\chi_{t}/2\pi$ } & {MHz } & 
{-6.3 } & {-10.7 }
\tabularnewline
{*transfer resonator/filter coupling, $J_{t}/2\pi$ } & {MHz } & {36.2 } & {36.4 }\tabularnewline
{*transfer Purcell filter bandwidth, $\kappa_{{\rm Pt}}/2\pi$ } & {MHz } & {166 } & {146 }\tabularnewline
{transfer resonator eff. bandwidth, $\kappa_{\rm A/B}/2\pi$ } & {MHz } & {26.7 } & {30.7 }\tabularnewline
\hline
\end{tabular}{\footnotesize{}\caption{\label{tab:ChipParameters} Measured device parameters for samples installed at nodes A and B, the parameters marked with (*) are taken from \cite{Storz2023}, see main text. The values of $\kappa_{\rm A/B}$ are extracted from calibrations of the f0-g1 transition, and may not perfectly match the other parameters of the transfer circuit, which were measured with transmission spectroscopy in \cite{Storz2023}.}


}
\end{table}

\section{Errors in the remote gate}
\label{app:errors}

The remote CPHASE gate realized in this work reaches a maximum fidelity of approximately $\mathcal{F} = 71\%$, see Fig.~{\ref{fig:CZ_DMs}}. The observed infidelity arises from a combination of photon loss, qubit dissipation, control imperfections, and limitations intrinsic to the process of photon scattering. In this appendix, we discuss the error contributions in detail.

The main source of error in this protocol is the loss of photons in the communication channel, which takes place predominantly in the microwave connections between the quantum devices and the waveguide. Each connection includes wire bonds, a gold-plated copper PCB, an SMA connector on the PCB, a NbTi coaxial cable and a coaxial-to-WR90 adapter. From independent measurements, we estimate these components to contribute 10 percentage points to the total photon loss per one-way trip of the photon \cite{Storz2023a}, which adds up to 20 percentage points in the bidirectional photon transfer. The 30-meter-long WR90 waveguide is superconducting and features measured attenuation below 1~dB/km, which makes its contribution negligible.

Another substantial source of incoherent error is dissipation in the emitter qubit, whose ef transition features a lifetime $T_1^{\rm ef} = 7.6~\upmu s$, longer than the duration $\tau$ of the photon emission operation only by 1.5 orders of magnitude. Assuming the excitation spends, on average, $\tau/2$ in the $\ket{\rm f}$ state during the $\ket{\rm f0}-\ket{g1}$ pulse, we estimate that the excitation is damped in the qubit by $2\times(1-e^{-\tau/2T_1^{\rm ef}})\approx3\%$ during the processes of emission and absorption. Additionally, photon loss is also affected by a design choice made for simplified characterization of the sample. The transfer resonators are weakly coupled to an input port used to measure their resonance frequency with a transmission measurement \cite{Storz2023}. The decay of the resonator into this port is expected to take about 1\% of the photon. In a bidirectional photon transfer protocol, where the photon populates the resonators in three occasions, this would add up to 3\% photon loss. The contributions discussed in this and previous paragraphs are independent of the frequency of the communication photon, and could be mitigated by using a more direct coupling between the quantum devices and the channel, namely with superconducting materials \cite{Qiu2025a, Niu2023}, and state-of-the-art fabrication methods \cite{Bland2025}.

In addition to frequency-independent loss and dissipation, the gate fidelity is affected by frequency-dependent control errors in the emission and absorption of shaped photons. These errors lead to a mismatch between the intended photon mode and the mode actually emitted into the communication channel. Such mismatches can arise from imperfect calibration of the emission resonator linewidths $\kappa_{\rm A/B}$, the magnitude of the AC-Stark shift (see App.\ref{app:f0g1_calibration}), the influence of a Purcell filter, or the approximate nature of the variable-frequency photon emission procedure itself (see Sec.~\ref{sec:Variable_frequency}). While these effects may not reduce the photon emission probability significantly (see Fig.~\ref{fig:f0g1_scheme} (d)), they can distort the amplitude and phase profile of the emitted wavepacket, reducing its overlap with the ideal absorption mode and therefore lowering the absorption fidelity. Further investigation is required to quantify the relative importance of these effects \cite{Miyamura2025, Pernas2026}.

\begin{figure}[hbtp]
    \centering
    \includegraphics[width=\linewidth]{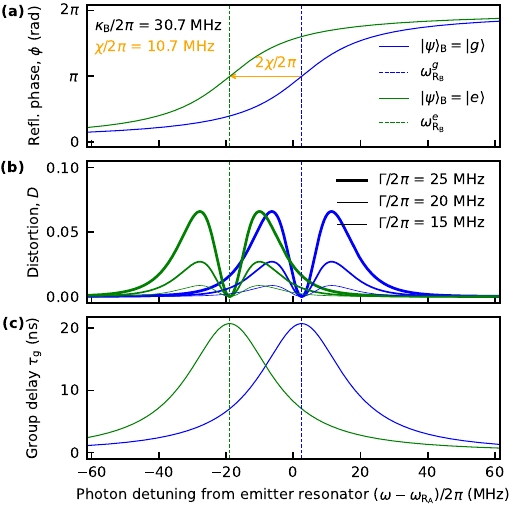}
    \caption{\textbf{Error mechanisms in the photon reflection}
    \textbf{(a)}~Phase acquired by the photon upon reflection at the remote resonator in node B.
    \textbf{(b)}~Distortion $D$ of the photon wavepacket upon reflection at the remote node, as a function of the detuning between the photon and the remote resonator. The color indicates the state of qubit B. Different line thicknesses indicate different photon bandwidths $\Gamma$.
    \textbf{(c)}~Group delay of the photon wavepacket vs frequency. The color indicates the state of qubit B.
    }
    \label{fig:app_distortion_and_delay}
\end{figure}

Finally, the remote CPHASE gate is subject to limitations intrinsic to the photon-scattering mechanism used to generate the conditional phase. The reflection coefficient in resonator B is not only nonlinearly dependent on the frequency of the photon but also on the state of qubit B due to the dispersive shift, see Eq.~(\ref{eqn:scatt_phi}), Fig.~\ref{fig:app_distortion_and_delay} (a) and Sec.~\ref{sec:Variable_frequency}. These two mechanisms yield a reduction in the efficiency of the photon reabsorption, constituting two sources of error in the operation. The nonlinear frequency dependency provokes a distortion $(f(t)\rightarrow f_{\rm refl}(t))$ in the photon envelope upon reflection, of magnitude $D = 1 - |\int f(t) f_{\rm refl}(t)dt|^2$, which we quantify following \cite{Penas2022} for both states of qubit B, see Fig.~\ref{fig:app_distortion_and_delay} (b). When the frequency equals that of the resonator for a given qubit state, or is far from the resonance by several times the linewidth, the distortion $D$ is minimal, as the photon experiences the linear-regime response of the resonator or does not enter the resonator at all, respectively. Conversely, when the photon-resonator detuning is nonzero but smaller or comparable to the resonator linewidth $\kappa_{\rm B}$, the phase becomes nonlinear and the wavepacket experiences a significant distortion $D$. This distortion becomes more detrimental as the photon bandwidth $\Gamma$ approaches $\kappa_{\rm B}$, see Fig.~\ref{fig:app_distortion_and_delay} (b). In addition, the state-dependence of the phase $\phi(\omega)$ is inherited by the group delay $\tau_g = d\phi/d\omega$, yielding a state-dependent optimal absorption time as explained in the main text. Only when this delay is the same for both states, a condition met at the frequency $(\omega_{\rm R_B}^g + \omega_{\rm R_B}^e)/2$ which corresponds to the maximal conditional phase $\Delta\phi=\pi$, the absorption efficiency is not affected by this mechanism. Any other conditional phase $\Delta\phi\neq\pi$ requires a different photon frequency, for which the optimal delay is state-dependent and the absorption efficiency decreases. This limitation is inherent to the presented gate scheme, and therefore its mitigation appears challenging.

\section{Calibration of the $\ket{f0}-\ket{g1}$ transition}
\label{app:f0g1_calibration}

\begin{figure}[hbtp]
    \centering
    \includegraphics[width = \linewidth]{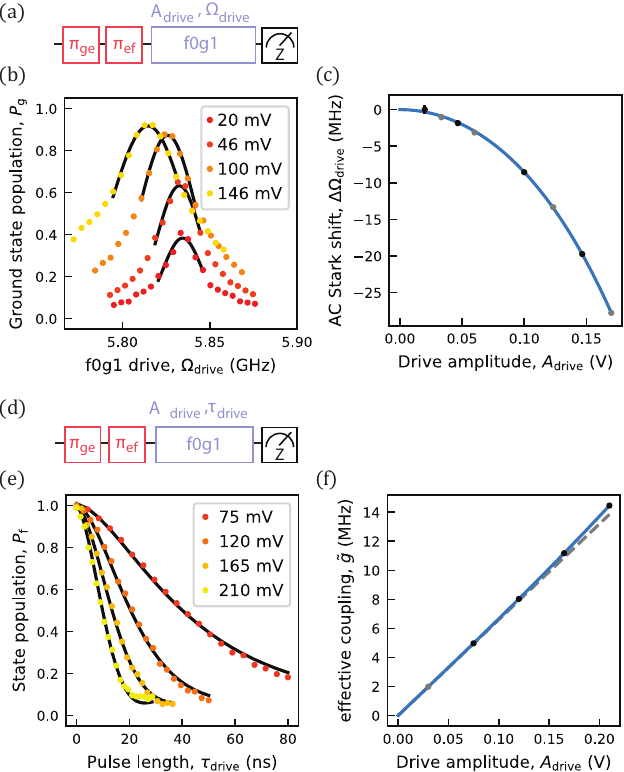}
    \caption{\textbf{(a)} Pulse scheme used for the AC Stark shift calibration of the $\ket{f0}$-$\ket{g1}$ transition. \textbf{(b)} Measured ground state population $P_g$ versus f0g1 drive frequency $\Omega_{\textrm{drive}}$ for different color-coded drive amplitudes $A_{\textrm{drive}}$, whose values are given in the inset legend. The solid lines are Gaussian fits to the measured data. \textbf{(c)} AC Stark shift of the $\ket{f0}$-$\ket{g1}$ transition as a function of the drive amplitude $A_{\textrm{drive}}$, with $\Omega_{\textrm{drive}}=5.834\,\textrm{GHz}$. Black markers with error bars are fitting results from (b), while gray markers denote results from additional measurements not displayed in (b). The solid line is an even fourth-order polynomial weighted fit.
    \textbf{(d)} Pulse scheme used for the Rabi rate calibration of the $\ket{f0}$-$\ket{g1}$ transition. \textbf{(e)} Measured second excited state population $P_f$ versus f0g1 drive pulse length $\tau_{\textrm{drive}}$ for different color-coded drive amplitudes $A_{\textrm{drive}}$, whose values are given in the inset legend. The solid lines are fits to the damped population dynamics described in the text. \textbf{(f)} Effective coupling rate $\tilde g$ of the $\ket{f0}$-$\ket{g1}$ transition as a function of the drive amplitude $A_{\textrm{drive}}$. Black markers are fitting results from (e), while gray markers denote results from additional measurements not displayed in (e). The solid line is an odd third-order polynomial fit; the gray line shows its linear contribution.
    }
    \label{fig:f0g1_calib}
\end{figure}

Using the $\ket{f0}$-$\ket{g1}$ transition for (variable-frequency) photon emission requires calibrating both the resonance condition, and the effective coupling strength induced by the microwave drive. In this appendix, we expand on the content of Fig.~\ref{fig:f0g1_scheme} and the calibration procedure outlined in the main text, and detail the measurements used to determine the AC Stark shift and the coupling rate $\tilde g$ as a function of the drive amplitude.

The resonant condition, as explained in the main text, takes the form $\Omega_{\textrm{drive}} = \omega_{ge} + \omega_{ef} - \omega_{\textrm{R}}$. However, since both qubit transitions, $\omega_{ge}$ and $\omega_{ef}$, are AC-Stark shifted depending on the amplitude of the drive pulse $A_{\textrm{drive}}$, we explicitly introduce the frequency shift term $\Delta\Omega_{\textrm{drive}}$, and calibrate its magnitude for a series of amplitudes. To do so, we prepare the qubit in the $\ket{f}$ state by applying $R_{ge}^\pi$ and $R_{ef}^\pi$ pulses, and follow with a flattop microwave pulse with amplitude $A_{\textrm{drive}}$, duration $t_{\textrm{f0g1}}$ and Gaussian\nobreakdash-shaped rise and fall edges, before performing qubit readout (see pulse pattern in Fig.~\ref{fig:f0g1_calib} (a)). For each drive amplitude $A_{\textrm{drive}}$, we sweep the frequency $\Omega_{\textrm{drive}}$ and record the resulting qubit population. The measurement is then repeated for a range of drive amplitudes. To keep the pulse area approximately constant, the pulse length is scaled inversely proportional to the pulse amplitude. The initial pulse area is chosen such that $P_g$ stays close to, yet below, unity.

The resonance condition corresponds to the maximum of the induced coupling between $\ket{f0}$ and $\ket{g1}$, resulting in the largest transfer of population from $\ket{f}$ to $\ket{g}$. Measured curves of $\ket{g}$-state population as a function of drive frequency for different drive amplitudes are shown in Figure~\ref{fig:f0g1_calib} (b). For larger $A_{\textrm{drive}}$ the resonance frequency moves downward, as expected for the AC Stark shift~\cite{Schuster2005} with the frequency configuration of the chip, see Appendix \ref{app:device}.

We extract the resonance frequency by fitting the measured ground\nobreakdash-state population $P_g$ with a Gaussian fit around the peak (black lines in Fig.~\ref{fig:f0g1_calib} (b)). Only data points exceeding a predefined population threshold are included in the fit; in the figure, this range is indicated by the endpoints of the fitting curves. The center frequency of the Gaussian fit is taken as the resonance frequency corresponding to the given drive amplitude.

Repeating this procedure for different drive amplitudes yields the AC Stark relation $\Delta\Omega_{\textrm{drive}}\leftrightarrow A_{\textrm{drive}}$. To obtain a continuous calibration curve, the extracted resonance frequencies are fitted with an even fourth-order polynomial $P(x)=a+bx^2+cx^4$. The resulting interpolation provides the frequency correction required for intermediate drive amplitudes and shows good agreement with the measured data, as illustrated in Fig.~\ref{fig:f0g1_calib} (c).

Having calibrated the AC Stark shift, we next determine the relation between the effective coupling strength $\tilde g$ and the drive amplitude $A_{\textrm{drive}}$. As before, we prepare the qubit in the $\ket{f}$ state, after which the emission pulse is applied. For a fixed drive amplitude, we vary the pulse length $\tau_{\textrm{drive}}$ (Fig.~\ref{fig:f0g1_calib} (d)), while setting the drive frequency according to the AC Stark shift calibration obtained above. The coupled qubit-resonator system undergoes Rabi oscillations between  $\ket{f0}$ and $\ket{g1}$, which are damped due to the spontaneous decay from $\ket{g1}$ to $\ket{g0}$~\cite{Magnard2018}.

We measure the second excited state population $P_{\textrm{f}}$, and plot it as a function of the length of the drive pulse $\tau_{\textrm{drive}}$ for different drive amplitudes $A_{\textrm{drive}}$. For most drive amplitudes, the population decays monotonically, indicating that the coupled qubit-resonator system is in the overdamped regime. As the drive amplitude increases, the population is transferred out of the $\ket{f}$ state more rapidly, indicating a larger induced coupling strength $\tilde{g}$ (see Figure~\ref{fig:f0g1_calib} (e)). At the largest drive amplitude ($A_{\textrm{drive}}=210\,\textrm{mV}$), the population $P_{\textrm{f}}$ starts to rise again for pulses $\tau\gtrsim20\,\textrm{ns}$, indicating that the system enters the under\nobreakdash-damped regime.

To model the measured population dynamics, we use the analytical expression derived in~\cite{Magnard2018},

\begin{equation}
P_{\textrm{f}(t)}=e^{-\frac{\kappa_{\rm A}+\gamma}{2}t}\left| \cosh{\frac{\Omega t}{2}}+\frac{\kappa_{\rm A}-\gamma}{2\Omega} \sinh{\frac{\Omega t}{2}} \right|^2,
\end{equation}
where $\kappa_{\rm A}$ is the effective decay rate of the resonator into the emission line, $\gamma$ is the rate of spontaneous decay of the qubit from $\ket{f}$ to $\ket{e}$, and $\Omega=\sqrt{-(2\tilde {g})^2+(\kappa_{\rm A}-\gamma)^2/4}$. If the drive rate $\tilde {g} > \left| \kappa_{\rm A} -\gamma\ \right| /4 $, the term $\Omega$ becomes imaginary, which yields underdamped oscillations of $P_{\textrm{f}}$, consistent with the behavior observed at the largest drive amplitude in Fig.~\ref{fig:f0g1_calib} (e). We expect that $\kappa_{\rm A}$ and $\gamma$, being fixed by the device geometry and the material properties of the chip, are independent of the drive amplitude. Therefore, these parameters are constrained to take the same values when fitting all datasets. The resulting fits, shown as solid lines in Fig.~\ref{fig:f0g1_calib} (e), show good agreement with the measured data.

As we did for the AC Stark relation, we use the extracted fit parameters to obtain a continuous calibration curve establishing the relation $A_{\textrm{drive}} \leftrightarrow \tilde{g}$. For weak drives, the effective coupling strength is expected to increase linearly with the amplitude~\cite{Zeytinoglu2015}, while higher-order effects can lead to a nonlinear dependence at larger amplitudes. We therefore fit the extracted values of $\tilde{g}$ with an odd third\nobreakdash-order polynomial, $P(x)=ax+bx^3$, shown as the solid line in Fig.~\ref{fig:f0g1_calib} (f). For comparison, the linear contribution $ax$ is displayed with a gray line.

The calibration for detuned photons is done by subtracting the desired photon detuning $\delta\omega$ from the calibrated AC Stark curve, as per step 2 of the box in the main text in Section~\ref{sec:Variable_frequency}. The measurement and analysis described in Figure~\ref{fig:f0g1_calib} (d-f) are then repeated using this shifted resonance condition.

\section{Optimal device parameters}

The conditional-phase gate presented in this work combines photon emission and absorption at the active node A with a dispersive interaction between resonator and qubit at the remote node B. This combination imposes a set of design requirements that must be taken into consideration to optimize the gate fidelity. Here we discuss these requirements using a single-pole model (see eq.~\eqref{eqn:scatt_phi}), and how they compete with each other.

In order to realize a conditional phase $\Delta \phi = \pi$ to achieve maximal entanglement between the two remote qubits, it is necessary that the dispersive shift between the transmon and the transfer resonator is at least half the linewidth of the resonator, that is $|\chi| \geq \kappa_{\rm B}/2$, see Fig.~\ref{fig:bidir_transfer} (c). As we want to preserve fast photon emission to avoid decoherence during the gate, the resonator linewidth $\kappa_{\rm B}$ must be large, so the dispersive shift $\chi = \alpha g_{\rm qt}^2/\Delta_{\rm qt}(\Delta_{\rm qt}+\alpha)$ \cite{Blais2021} must be large too. For this purpose, one must realize a large qubit anharmonicity $\alpha$, a large qubit-resonator coupling strength $g_{\rm qt}$ and a moderate qubit-resonator detuning $\Delta_{\rm qt}$.

However, an arbitrarily large dispersive shift $\chi$ would deteriorate the gate fidelity due to a different effect. When $|\chi|=\kappa_{\rm B}/2$, the phase $\Delta \phi = \pi$ requires the emission of a photon detuned from the remote resonator by $\omega_{\rm ph} - \omega_{\rm R_B} = \chi$. At this precise frequency, the optimal absorption delay is the same independently of the state of the remote qubit, see Fig.~\ref{fig:phase_tunability} (d) and \ref{fig:app_distortion_and_delay} (c). Nonetheless, when $|\chi|>\kappa_{\rm B}/2$, the maximal conditional phase requires a photon detuning smaller than $\chi$. With this parameter choice, the optimal delay becomes dependent on the state of the remote qubit, yielding a suboptimal absorption fidelity. For this reason, the single-pole model suggests the optimal value for the dispersive shift of $|\chi|=\kappa_{\rm B}/2$, a condition that has also been found optimal to maximize signal-to-noise ratio in readout schemes which, as our entangling gate, leverage the dispersive interaction between a transmon qubit and a resonator \cite{Walter2017}. 

In the present devices, the static dispersive shifts extracted from spectroscopy are of the same order as the transfer-resonator linewidths, placing the devices close enough to this regime to demonstrate a large conditional phase. More specifically, the ratio for the node B is $\chi_{\rm B}/\kappa_{\rm B} = -0.35$, which predicts a slightly lower conditional phase than the one we observe experimentally. Quantitative agreement with the measured phase response, however, is affected by the Purcell-filter hybridization and by the effective linewidth seen by the finite-bandwidth communication photon.

Another difference in the requirements for this remote gate as compared to a state transfer operation \cite{Kurpiers2018} is the optimal detuning between the transfer resonators at each node. State transfer benefits from having no frequency difference between these resonators: $\Delta_{\rm AB} = \omega_{\rm R_A}-\omega_{\rm R_B} = 0$, as this relaxes the need for variable-frequency photon emission \cite{Miyamura2025}. Nevertheless, as mentioned above, for the dispersive interaction in the presented remote gate to induce a maximal phase contrast $\Delta\phi = \pi$, the photon must be detuned from the remote resonator by exactly the dispersive shift $\chi$. Therefore, the configuration where the need for variable-frequency photon emission is relaxed is $\Delta_{\rm AB}=\chi$ instead.

\section{Repeated remote readout and classical error correction}
\label{app:remote_readout}

\begin{figure}[hbtp]
    \centering
    \includegraphics[width=\linewidth]{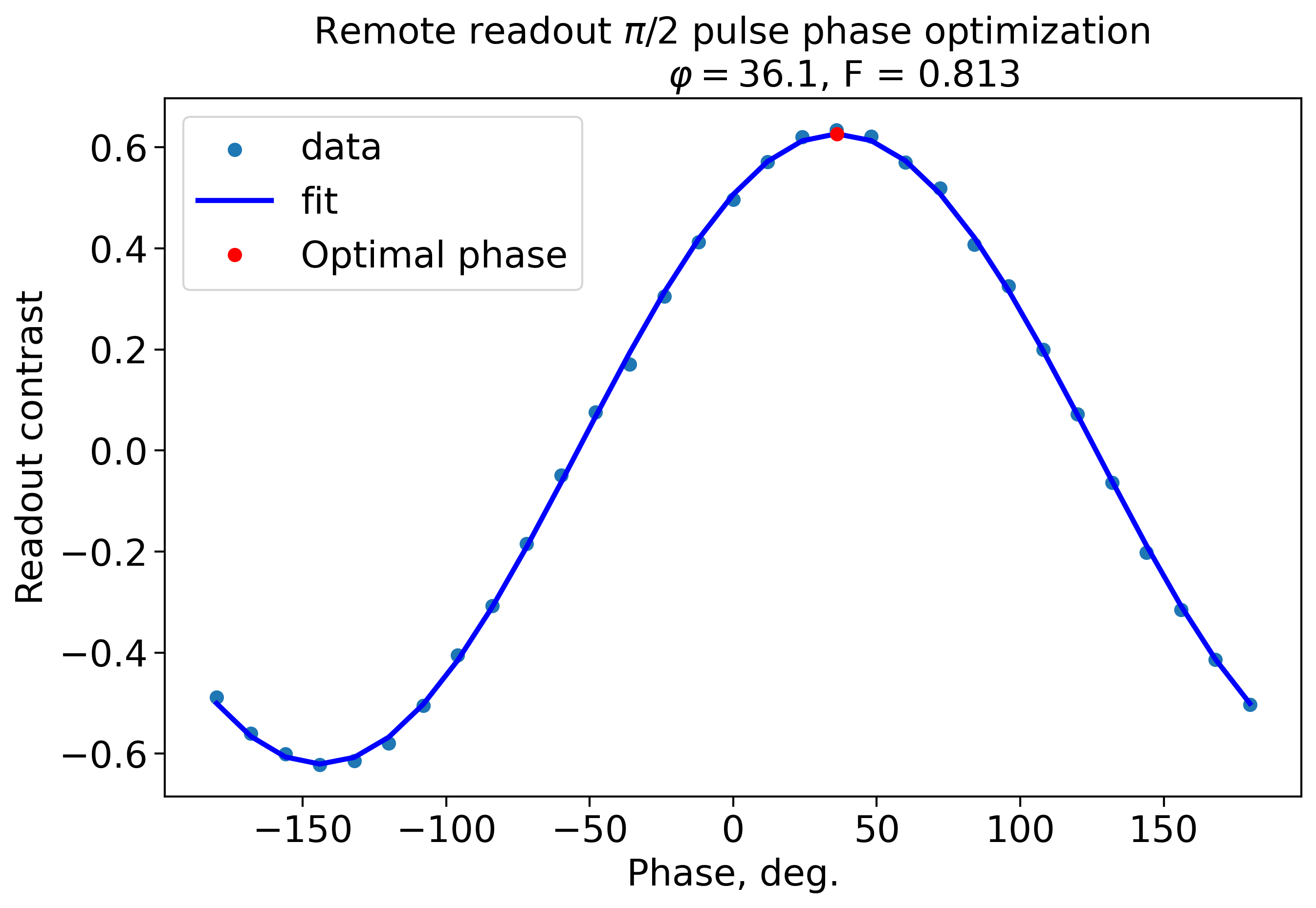}
    \caption{\textbf{Optimization of the final $\pi/2$-pulse phase used in the remote readout protocol.} 
    Difference between the measured excited-state populations for the remote qubit prepared in the $\ket{g}$ and $\ket{e}$ states as a function of the phase of the final $\pi/2$ pulse. The markers are measured data and the solid curve is the cosine fit. The red circle marks the optimal phase of $36.1$~deg.
    }
    \label{fig:app_remote_readout_phase}
\end{figure}

The remote readout protocol requires calibrating the phase of the final $(\pi/2)$ pulse applied to the emitter qubit. To determine the optimal phase, we prepare the remote qubit in either the $\ket{g}$ or $\ket{e}$ state and perform the complete remote readout sequence while sweeping the phase of the final rotation pulse. For each phase value, we evaluate the remote readout contrast, defined as the difference between the measured excited-state populations of the emitter qubit obtained for the two remote-qubit preparations. The resulting phase dependence follows a sinusoidal behavior, from which we extract the phase maximizing the readout contrast using a cosine fit. This phase is unconditional and should not be confused with the conditional phase acquired by the communication photon during the remote controlled-phase gate. Rather, it defines the measurement basis of the emitter qubit and maximizes the sensitivity of the remote readout operation to the state of the remote qubit. The phase sweep and the corresponding cosine fit are shown in Fig.~\ref{fig:app_remote_readout_phase}. Once calibrated, the phase remains fixed for all subsequent remote readout experiments, and the remote readout protocol can be used standalone as a standard measurement operation, or repeatedly as we show next.

After calibrating the phase of the final $(\pi/2)$ pulse, the remote readout protocol can be repeated multiple times in succession. To realize repeated remote readout, we concatenate several instances of the remote readout sequence, measuring the emitter qubit after each repetition. Following each measurement, we introduce a 100~ns delay time to allow for depletion of the readout resonator before the next repetition begins.

\begin{figure} [t]
    \centering
    \includegraphics[width=\linewidth]{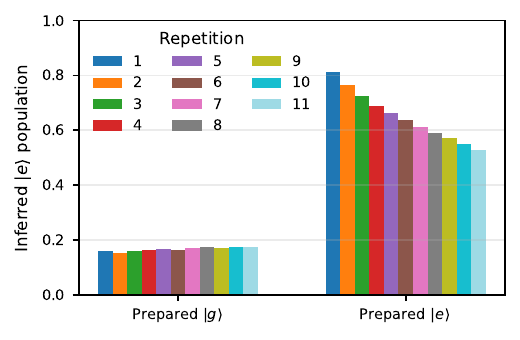}
    \caption{\textbf{Inferred population of the remote qubit as a function of the repetition number.} 
    The remote qubit is prepared either in the $\ket{g} \mbox{ or }\ket{e}$ state, and the remote readout protocol is repeated eleven times in succession. The bars show the inferred excited-state population of the remote qubit obtained from each individual repetition of the protocol. The two groups of bars correspond to the two prepared states, and the bar colors indicate the repetition number.
    }
    \label{fig:app_repeated_RR_individual}
\end{figure}

As discussed in the main text, the inferred remote state is obtained from the sum of the emitter-qubit preparation state and the local readout result modulo two. The resulting inferred excited-state populations are shown in Fig.~\ref{fig:app_repeated_RR_individual}. For a remote qubit prepared in $\ket{g}$, the inferred population remains approximately constant throughout the measurement sequence. For a remote qubit prepared in $\ket{e}$, the inferred population gradually decreases with repetition number, reflecting energy relaxation of the remote qubit during the repeated measurement process. The observed behavior is consistent with the observation that the dominant source of readout errors is photon loss during the remote readout process. Indeed, with the photon loss of the bidirectional photon transfer protocol close to 30\%, the ground state is misidentified slightly more than 15\% of the time, as the photon loss leads to a random outcome. Since photon-loss events are largely independent between repetitions, successive executions of the protocol provide multiple nearly uncorrelated estimates of the state of the remote qubit. This enables the use of classical error-correction techniques, such as majority voting and weighted decoding, to substantially improve the overall readout fidelity.

\begin{figure}[hbtp]
    \centering
    \includegraphics[width=\linewidth]{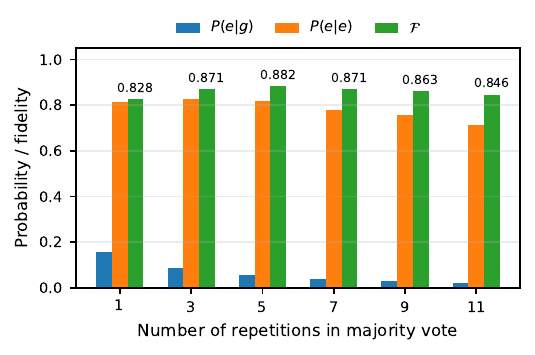}
    \caption{\textbf{Classical error correction of repeated remote readout using majority voting.}
    Assignment probabilities and corresponding readout fidelity obtained by combining different odd numbers of repeated remote readout outcomes. For each experimental run, the inferred remote-qubit state is determined by majority vote over the indicated number of repetitions. Blue bars show the probability $\mbox{P}(e|g)$ to assign the remote qubit to $\ket{e}$ when it was prepared in $\ket{g}$, orange bars show the probability $\mbox{P}(e|e)$ to assign $\ket{e}$ when it was prepared in $\ket{e}$, and green bars show the resulting readout fidelity $\mathcal{F}$. The fidelity reaches its maximum for five repetitions and decreases for longer voting windows.
    }
    \label{fig:app_repeated_error_correction}
\end{figure}

We first perform classical error correction on a repeated remote readout data using simple majority voting over an odd number of repetitions. For example, when using seven repetitions, readouts 1-7 are combined to determine a single inferred state of the remote qubit for each experimental run. The error-corrected readout fidelity is shown in Fig.~\ref{fig:app_repeated_error_correction} as a function of the number of repetitions included in the vote. The fidelity increases from the single-shot value of 82.8\% to a maximum value of 88.2\% for five repetitions, demonstrating the efficiency of the method and being consistent with different readout attempts providing largely independent information about the state of the remote qubit. For longer sequences, the fidelity decreases again as energy relaxation of the remote qubit reduces the information gained from later repetitions.

A further improvement can be obtained by replacing majority voting with a weighted sum of the individual readout outcomes followed by thresholding. The weighted decoding procedure is directly inherited from optimal integration of dispersive readout traces. Each repetition is assigned a weight proportional to its measured discrimination contrast (i.e. $w_i = P_e^i - P_g^i)$, such that repetitions carrying more information about the state of the remote qubit are weighted more strongly. We then determine the optimal discrimination threshold numerically by maximizing the resulting readout fidelity. Using this procedure, we obtain a maximum remote readout fidelity of 89.4\% for seven repetitions, as reported in panel \textbf{c} of the Fig.~\ref{fig:Remote_readout} in the main text.

\end{document}